\documentclass{iaus}
\usepackage{graphicx}

\title[C-rich molecules in space] 
{A review on carbon-rich molecules in space}

\author[Cataldo, Garc\'{\i}a-Hern\'andez \& Manchado]   
{Franco Cataldo$^{1,2}$, D. A. Garc\'{\i}a-Hern\'andez$^{3,4}$ 
\and Arturo Manchado$^{3,4,5}$}

\affiliation{$^1$INAF - Osservatorio Astrofisica di
Catania, Via S. Sofia 78, 95123, Italy\\[\affilskip]
$^2$Actinium Chemical Research, Via Casilina 1626/A, 00133 Rome, Italy\\[\affilskip]
$^3$Instituto de Astrof\'{\i}sica de Canarias, V\'{\i}a Lactea s/n, E-38205, La Laguna, Tenerife, Spain\\[\affilskip]
$^4$Universidad de La Laguna (ULL), Dept. de Astrof\'{\i}sica, E-38205, La Laguna, Tenerife, Spain
$^5$CSIC, Madrid, Spain}

\pubyear{2012}
\volume{xxx}  
\pagerange{119--126}
\jname{Unexplained Spectral Phenomena in the ISM}
\editors{S. Kwok ed.}
\begin{document}

\maketitle

\begin{abstract}
We present and discuss carbon-rich compounds of astrochemical interest such as
polyynes, acetylenic carbon chains and the related derivative known as
monocyanopolyynes and dicyanopolyynes. Fullerenes are now known to be abundant
in space, while fulleranes - the hydrogenated fullerenes - and other carbon-rich
compounds such as very large polycyclic aromatic hydrocarbons (PAHs) and heavy
petroleum fractions are suspected to be present in space. We review the
synthesis, the infrared spectra as well as the electronic absorption spectra of
these four classes of carbon-rich molecules. The existence or possible existence
in space of the latter molecules is reported and discussed.
\end{abstract}

The review regards a series of carbon-rich molecules which were recently studied
by the authors. First we will discuss the first class of carbon-rich molecules:
the polyynes and the related molecules known as monocyanopolyynes and
dicyanopolyynes. These molecules were detected by radioastronomy in carbon-rich
Asymptotic Giant Branch (AGB) stars, in dense and dark interstellar clouds like
TMC-1 and in hot molecular cores. Although polyynes are highly unstable in
terrestrial conditions, Cataldo (2004, 2006a,b) has developed a synthesis of
these molecules by using a carbon arc and trapping the molecules in a solvent.
When the arc is struck in liquid hydrocarbons, all the polyynes homologue series
from H-(C$\equiv$C)$_{3}$-H to H-(C$\equiv$C)$_{9}$-H is obtained and the electronic
absorption spectrum of each individual polyyne molecular specie has been
recorded. The formation of polyynes is always accompanied by the formation of
PAHs and carbon soot. Indeed, polyynes are considered the precursors of PAHs and
carbon soot. When the carbon arc is struck in liquid ammonia or in acetonitrile,
then a mixture of polyynes series and monocyanopolyynes series
H-(C$\equiv$C)$_{n}$-CN is obtained. Dicyanopolyynes NC-(C$\equiv$C)$_{n}$-CN can be
synthesized by striking the carbon arc in liquid nitrogen. It was proposed that
the laboratory conditions of polyynes synthesis through the carbon arc are
comparable to those existing in the circumstellar medium of certain AGB stars
since both in laboratory conditions and in the circumstellar environment the
mechanism of carbon chain formation follows a free radical path (Cataldo
2006a,b). More recent work has shown that polyynes are also formed in the gas
phase by arcing graphite in a flow of argon, in which the addition of methane to
Ar strongly enhances the formation of polyynes, which are then accompanied by
PAHs such as naphthalene (Cataldo 2007). Polyynes are an endothermal compound
with a positive free energy of formation from the elements at 298 K. However,
their Gibbs free energy of formation becomes negative just above 4400 K, which
is the temperature of the carbon arc. The polyynes are thermodynamically stable
and form quite easily at such high temperatures, but are then quenched to a
lower temperature by the hydrocarbon solvent surrounding the arc and become
trapped there, thus permitting their manipulation, separation and analysis
(Cataldo 2007). Another interesting carbon-rich class of molecules are
fullerenes, which were recently detected in a series of astrophysical objects
(Cami et al. 2010; Garc\'{\i}a-Hern\'andez et al. 2010, 2011a,b; Sellgren et al. 2010;
Zhang \& Kwok 2011). Fullerenes show considerable stability toward the action of
high energy radiation so that they could survive for billions of years in the
ISM under certain conditions (Cataldo et al. 2009). This fact may explain why
fullerenes have been found in completely different astrophysical objects. To
facilitate the search of C$_{60}$ and C$_{70}$ fullerenes in space, we have studied the
dependence of infrared band pattern and band intensity of these molecules with
temperature and measured the molar absorptivity for the quantitative
determination of their abundance (Iglesias-Groth et al. 2011). Furthermore, the
electronic absorption spectra of the radical cation of C$_{60}$ and C$_{70}$ in a very
high dielectric constant medium has been determined (Cataldo et al. 2012).
Fullerenes are very reactive with atomic hydrogen, and form fulleranes, which
are hydrogenated fullerenes (Cataldo \& Iglesias-Groth 2010). When heated or
photolyzed, the fulleranes release molecular hydrogen yielding back the parent
fullerene. Therefore, fullerenes may play a key role in the molecular hydrogen
formation starting from atomic hydrogen. The photolysis rate constant of
fulleranes appears of the same order of magnitude as that of polyynes (Cataldo \&
Iglesias-Groth 2009). Curiously, the fullerane C$_{60}$H$_{36}$ shows an electronic
absorption spectrum with a unique peak at 217 nm, exactly matching the UV ``bump"
of the interstellar light extinction curve (Cataldo \& Iglesias-Groth 2009). The
infrared spectra of a series of reference fulleranes have been recorded in the
laboratory by Iglesias-Groth et al. (2012) together with the relative molar
absorptivity. Consequently, since the reference spectra are now available, a
search for fulleranes can now made in space, and their possible detection in
astrophysical environments is now only a matter of time and luck. Coal was
proposed by Papoular et al. (1989) as a possible model for matching the
infrared band pattern of the unidentified infrared bands (UIBs) of certain
planetary nebula (PNe) and  proto-PNe (PPNe). However, anthracite, "mature"
coal, is eminently aromatic with minor aliphatic components. More recently,
Cataldo et al. (2003, 2004a,b) have shown that certain heavy petroleum fractions
are also able to match the band pattern of coal and of PPNe and PNe. Thus,
petroleum fractions should be used as realistic model compounds since they are
composed by a "core" of 3 to 4 condensed aromatic rings surrounded by
cycloaliphatic (naphthenic rings) and chains of aliphatic sp3 hybridized carbon.
Thus, instead of searching for pure PAHs, the heavy petroleum fraction model
offers molecules where the aromatic, naphthenic and aliphatic moieties co-exist,
and match certain PPNe spectra where a mixture of aliphatic/cycloaliphatic and
aromatic structures are evident. Very large PAHs (VLPAHs) are not easily
accessible but of high interest as reference molecules for the explanation of
the diffuse interstellar bands (DIBs) of the interstellar medium (ISM). Using
the Scholl reaction, we have synthesized a series of VLPAHs ranging from
dicoronylene to quaterrylene to hexabenzocoronene. Dicoronylene was also
obtained by the thermal dimerization of coronene. If the thermal treatment of
coronene is prolonged, the oligomerization of coronene proceeds further,
yielding a black oligomer, which is probably a trimer or a higher homologue. It
is shown that from coronene oligomerization it is possible to build a sheet of
graphene.


\end{document}